\documentclass[11pt]{article}

\usepackage[]{graphics,graphicx}

\usepackage{amsmath}
\usepackage{pstricks}

\usepackage{blindtext}

\usepackage{setspace} 

\RequirePackage{lineno}

\begin{document} 
\modulolinenumbers[1]


\begin{center}
{\LARGE\bf The landslide drag}
\end{center}
\vspace{7mm}
{Shiva P. Pudasaini
\\[3mm]
Technical University of Munich, 
School of Engineering and Design\\
Civil and Environmental Engineering}\\
{Arcisstrasse 21, D-80333, Munich, Germany}\\[3mm]
Kathmandu Institute of Complex Flows\\ Kageshwori Manohara - 3,
Bhadrabas, Kathmandu, Nepal\\[2mm]
{E-mail: shiva.pudasaini@tum.de}\\[7mm]
\noindent
\noindent
{\bf Abstract:}
{Drag is one of the most important energy dissipation mechanisms in nature, including landslides and debris flows. To satisfactorily reproduce laboratory or field data in simulating landslides, often empirical relations or convenient numerical values are used for the drag force coefficient. However, this is just a parameter calibration rather than a physical reality. Why should the drag coefficient be a constant for a dynamically evolving landslide? Which drag coefficient represents the physical reality? So, what exactly is the drag remains an open question. As the landslide is a deformable body, the drag-deformation-flow must be interconnected. Empirical drag coefficients lack important dynamical aspects. As the drag coefficient is less likely to be measurable, it must be described with some mechanical models. Yet, there exists no analytical model for the drag coefficient. Here, we postulate that the drag coefficient must be a function of the evolving landslide velocity, as it must contain information constituting the landslide acceleration in relation to the net driving acceleration. We develop an innovative, evolutionary drag coefficient that adjusts automatically during the landslide motion as it includes the viscosity of the ambient fluid. The drag coefficient is described by a dimensionless acceleration number as it is regulated by the physics and dynamics of the flow. Formal derivation shows that the drag coefficient is the measure of energy inefficiency. This settles down the deliberation on the drag force in landslide dynamics, reshaping the concept of drag. Simulation results highlight the essence, mechanical strength and functionality of the proposed analytical drag as it demonstrates the inherent frictional behaviour of granular debris flows. As the dynamical drag coefficients appeared to be around the often calibrated values, the new drag potentially well reproduces natural event dynamics, but now with clear physical basis.}

\section{Introduction}

Drag is a fundamental force in nature (Hill, 1928; Hoerner, 1965; Kroo, 2001; White, 2016; Anderson, 2023). It is induced when an object moves relative to its ambient fluid. The drag (also called the viscous drag, or the ambient drag) coefficient plays the central role in the drag force.
Drag is one of the principle energy dissipation mechanisms in landslide as it plays the dominant role in controlling the landslide dynamics, its run-out and deposition morphology (Voellmy, 1955; Salm 1966; Perla et al., 1980; Kelfoun, 2011; Salmanidou et al., 2018; Franz et al., 2021; Trujillo-Vela et al., 2021; Tayyebi et al., 2022; Singh et al., 2023).
However, there exists no explicit mechanical, analytical model for the drag force coefficient. Yet, in practice, some empirical relations and/or convenient numerical values are used for the drag coefficient in simulating the landslide dynamics without any physical foundation (Kattel et al., 2016; Pudasaini and Mergili, 2019; Pudasaini and Krautblatter, 2022), but, based on the model calibration (Brufau et al., 2000; Chen et al., 2006; Hungr and McDougall, 2009; Christen et al., 2010; Frank et al., 2015; Melo et al., 2018; Shugar et al., 2021;  Meyrat et al., 2023; Mergili et al., 2025; Sattar et al., 2025). Pudasaini (2024) presented a drag model, which, however, is limited to the particle moving with the surface wave speed, but not containing the physical and geometrical parameters, and not covering the flow dynamics.
\\[3mm]
 Here, we are concerned about the drag force coefficient for a complex deformable body such as landslides and avalanches (Shugar et al., 2021; Pandey et al., 2024; Mergili et al., 2025; Sattar et al., 2025). Till the date, the drag coefficient is largely determined experimentally or numerically (Voellmy, 1955; Christen et al., 2010; Melo et al., 2018; Zhang et al., 2021; Byers et al., 2024; Munch et al., 2024; Pudasaini and Mergili, 2025; Zhuang et al., 2025; Feng et al., 2026). However, from the dynamical perspective, the state of our knowledge on drag is thin: dynamically what exactly is the drag remains an open question. We postulate: the drag coefficient should be perceived as an evolutionary quantity rather than an empirical constant as considered classically. There are several reasons for why we need a dynamically explained drag coefficient for landslide motion (in general, for any complex mass transport problem) rather than just some empirically selected constant values. The motivation and need for a new evolutionary drag model lies in the following intuitions and facts.

\begin{itemize}
\item The landslide velocity changes, i.e., it accelerates as the mass is released, it may attain and retain a type of steady-state for some while as the mass moves in the main track, and decelerates and finally comes to a halt in the run-out (Wang et al., 2023; Mergili et al., 2025; Pudasaini and Mergili, 2025; Rinzin et al., 2025; Feng et al., 2026). Moreover, each differential part of the landslide moves with different velocity. From this perspective, it is  natural to think that the drag coefficient and the landside motion must somehow be interconnected. So, the drag coefficient must evolve throughout the landslide body, over time, and at each location of its propagation path.  In another words, effectively, the drag coefficient must be a function of the dynamically evolving landside velocity rather than just an empirical coefficient chosen globally for the entire landslide path and for the entire time of motion, from inception, through the track, to the final deposition.

\item The drag coefficient must contain the information about the track geometry, the material composition (i.e., the amount of solid particle and the fluid in the mixture), and the material parameters of the landslide, namely, the particle and fluid densities and friction. These are essential details that constitute the net driving force (Pudasaini and Krautblatter, 2022) determining the landslide motion that should be connected with the drag.

\item The drag coefficient must also depend on the actual acceleration of the propagating mass in relation to the net driving acceleration.

\item In aerodynamics and hydrodynamics, often, the shape of the object remains unchanged (Hill 1928; Hoerner, 1965; Hernandez Gomez et al., 2013). So, the drag (or, the resistance) coefficient may remain constant. However, the shape of a landslide changes continuously, sometimes very rapidly (Singh et al., 2023; Wang et al., 2023; Munch et al., 2024; Mergili et al., 2025; Rinzin et al., 2025; Feng et al., 2026). This is of particular interest for us. This includes the situations: as the mass is released, as it impacts some obstacles, when the topographic changes are rapid, and when the mass begins to deposit.

\item  In contrast to the solid body, for deformable body like landslides, the drag must be solved as a coupled problem as the drag-deformation-flow are intrinsically interconnected. This is so, because, as the fluid pushes the body, the body changes its shape. The new shape leads to changes in the flow. This again results in changes in the drag. In these dynamically unfolding situations, one cannot so easily measure or consider a single effective frontal area and the surface area in contact with the ambient fluid. Examples include: the flow of rock avalanche (Pudasaini et al., 2024), rock-ice avalanche (Shugar et al., 2021; Munch et al., 2024; Pudasaini, 2025b) and the glacial lake outburst flows (Kattel et al., 2016; Sattar et al., 2025). Empirical drag coefficients lack all these important physical, mechanical and dynamical aspects of the landslide motion, and its connection with the drag force. It appears that, for complex deformable body (Christen et al., 2010; Melo et al., 2018;  Shugar et al., 2021; Zhang et al., 2021; Byers et al., 2024; Munch et al., 2024; Pandey et al., 2024; Mergili et al., 2025; Sattar et al., 2025; Zhuang et al., 2025; Feng et al., 2026), the drag coefficient is less likely to be measurable. It must thus be described with some mechanical-dynamical models.
\end{itemize}
So, from the physical stand, the empirical drags cannot genuinely describe the complex nature of the drag force in landslide dynamics. For this, we need a physically described drag model. Here, we develop a well-defined, novel drag model. Guided by the physical-material parameters, the topography, the mechanics and the flow dynamics, this innovative, evolutionary drag coefficient adjusts automatically during the landslide motion. The drag dynamics is explained with a single dimensionless number, called the acceleration number, the first of its kind, which is the ratio between the flow acceleration and the net driving acceleration. This makes the model unique as it defines the strength of the analytical-dynamical drag. Based on the state of the  acceleration number, the model determines the mechanically evolving drag coefficient during the landslide motion.
The formally derived drag coefficient appears to be the measure of landslide (energy or mobility) inefficiency {as it takes into account the viscosity of the ambient fluid}. This addresses the long-standing problem in landside dynamics, reshaping the concept of the drag force in complex mass flow situations.
The model is applied for landslide motion down a channel. Simulation results demonstrate the mechanism and the functionality of the novel analytical drag model and its need in real event simulation.

\section{Different drags and their representations}

\subsection{The form and skin drag}

The drag force is induced when an object moves through a fluid (Hill, 1928; Hoerner, 1965; Kroo, 2001; Hernandez Gomez et al., 2013). There are two types of drags we are interested in: the form drag (or, the pressure drag) and the surface drag (or, the skin friction drag).
They happen for different physical reasons (White, 2016; Anderson, 2023). The form drag is caused by the shape of the moving object, and how it redistributes the fluid flow around it, and is proportional to projected cross-sectional area of the object perpendicular to the fluid flow. It does not explicitly contain the viscosity of the ambient fluid, however,
the viscosity still affects the form drag indirectly through the drag coefficient. The skin drag is caused by the fluid viscosity, the fluid rubbing against the surface of the object.
The force associated with the skin friction is derived from fluid shear stress integrated over the entire wetted surface area of the object in contact with the fluid.

\subsection{The total drag}

In practical and engineering applications, with adjusted projected cross-sectional area $A$, the form ($C_p$) and skin ($C_f$) drag coefficients are combined into a single total drag coefficient, $C_D$: $C_D = C_D\left(C_{p}, C_{f}\right)$. The fluid viscosity is implicit in $C_D$. Then, the total drag force $F_D$ can be written as (Hill, 1928; Hoerner, 1965; Kroo, 2001; Hernandez Gomez et al., 2013; White, 2016; Anderson, 2023):
\begin{equation}
\displaystyle{F_D = \frac{1}{2} C_D A \rho_a u^2},
\label{Eqn_a01}
\end{equation}
where, $u$ is the velocity of the object in motion and $\rho_a$ is the density of the ambient fluid.
$C_D$ is the weighted average of the form drag and the skin drag coefficients in the following way. The pressure component describes how much the shape blocks the fluid (or mobilizes, depending on what is moving, the fluid or the object). The skin friction component models how much the surface of the moving body in contact with the fluid applies drag on to it. This scales how large the skin friction is compared to the face induced pressure drag.

\subsection{The compact drag representation}

Assembling everything into a single entity $\beta$ (also called the ambient or the viscous drag coefficient), (\ref{Eqn_a01}) reduces to (Pudasaini and Fischer, 2020a):
\begin{equation}
\displaystyle{F_\beta = \beta u^2},
\label{Eqn_a02}
\end{equation}
where, $\displaystyle{}\beta$ in this formulation consistently has the dimension of 1/[m], that, with proper modelling arrangement, can be realized to represent $C_D A \rho_a/2$.
So, structurally, $\beta$ changes if the factors constituting the form and skin drag change, i.e., the evolving form, the physical parameters and the material properties of the moving body, the basal topography, and the ambient fluid. With this simple representation, we obtain the total resistance from the ambient fluid mass.
\\[3mm]
The compact representation for the drag force $\beta u^2$ in (\ref{Eqn_a02}) appears to be very effective in geophysical mass flow dynamics for deformable bodies, such as avalanches, landslides and debris flows, than the standard aerodynamic and hydrodynamic drags applied to rigid objects. This also simplifies the mathematical description. By using the expression  as in (\ref{Eqn_a02}) Pudasaini and Fischer (2020a) and Pudasaini and Krautblatter (2022) bundle the geometry and the drag coefficient into the single drag entity $\beta$. In this form, by using some simple calculus, one can quickly solve the momentum equation analytically for the velocity of the object in motion. Here, we follow this simple structure of the drag force in constructing the evolutionary, analytical drag coefficient.
\\[3mm]
In the depth-averaged frame, the final drag force is given by $\beta h u^2$ (Pudasaini and Fischer, 2020a; Pudasaini and Krautblatter, 2022) that is employed in the momentum balance equation at Section 4. So, here, the flow depth (per unit width) plays the role of the projected area of the moving body. From this mechanical perspective, the Voellmy-type models (Christen et al. 2010; Frank et al., 2015) are physically inconsistent as the drag coefficient in those models do not involve the flow depth $h$, one of the main dynamical quantities in the depth-averaged model. Moreover, evidently, in the Voellmy-type model, the definition of the ``turbulent'' coefficient does not fit into the frame of the viscous drag representation.

\section{The evolutionary mechanical drag model}

\subsection{The landslide velocity}

Consider the landslide velocity model (Pudasaini and Krautblatter, 2022):
\begin{equation}
\frac{du}{dt} = \alpha - \beta u^2,
\label{Eqn_1}
\end{equation}
where, $t$ is time, $u$ is the landslide velocity along the slope, $\alpha$ is the net driving force that includes: (i) the slope parallel gravity acceleration and the Coulomb frictional resistance:
\begin{equation}
\alpha = g\sin(\zeta) - (1-\gamma)\alpha_s g \cos(\zeta)\mu,
\label{Eqn_a}
\end{equation}
where, $g$ is the acceleration due to gravity, $\zeta$ is the slope angle, $\gamma = \rho_f/\rho_s$ is the ratio between the fluid and the solid density, $\alpha_s$ is the solid volume fraction in the landslide mixture material, and $\mu = \tan(\delta)$ is the friction coefficient with $\delta$ the friction angle of the solid particles,
and (ii) $\beta$ the coefficient of drag (resistance).
For $\alpha > 0$, the analytical solution of (\ref{Eqn_1}) is given by (Pudasaini and Krautblatter, 2022):
\begin{equation}
u = \sqrt{\frac{\alpha}{\beta}}\tanh\left[\sqrt{\alpha\beta}~t\right],
\label{Eqn_2}
\end{equation}
which will be employed to construct the dynamical drag model.

\subsection{The dynamical drag equation for positive net driving acceleration: $\alpha > 0$}

As we are mainly concerned in developing a model for the drag coefficient $\beta$, we reformulate (\ref{Eqn_2}) in terms of $\beta$ as follows:
\begin{equation}
\left(\frac{u}{\sqrt{\alpha}}\right)\sqrt{\beta} = 
\tanh\left[\left(\sqrt{\alpha}~t\right)\sqrt{\beta}\right].
\label{Eqn_3}
\end{equation}
With the definitions: $\displaystyle{S = \frac{u}{t}\frac{1}{\alpha} = \frac{u/t}{\alpha} = \frac{a}{\alpha}}$, which is independent of $\beta$, the main entity here, and $\eta = \sqrt{\alpha\beta}~t$, (\ref{Eqn_3}) can be re-written in a compact form as:
\begin{equation}
S\eta = \tanh(\eta).
\label{Eqn_4}
\end{equation}
Physically, when the tangent hyperbolic function is applied to the non-dimensional quantity $\eta$, it represents the authentic velocity of the landslide. This shows that, as $a = u/t$ can be realized as the (momental) dynamic acceleration of the flow, $S$ is a dimensionless quantity as the ratio between the flow acceleration $a$ and the net driving acceleration $\alpha$.
As invented here, we call $S$ the acceleration number. As we will see later, this determines everything.
However, note that, whether the acceleration number increases or decreases depends on how the velocity changes as time proceeds. That dynamics is also controlled by the applied net driving acceleration, $\alpha$.
Equation (\ref{Eqn_4}) can be recast in the equivalent functional form:
\begin{equation}
F(S; \eta) \equiv S \eta - \tanh(\eta) = 0.
\label{Eqn_5}
\end{equation}
We call (\ref{Eqn_5}) the dynamical drag model, or the $S$ function. This constitutes the following proposition:
\\[3mm]
{\it {\bf Proposition 1:} For each acceleration number $S$, there exists a unique positive number $\eta$ satisfying (\ref{Eqn_5}) as its root. Then, the drag coefficient is given by
\begin{equation}
\beta = \left({\eta}/{t} \right)^2/\alpha,
\label{Eqn_6}
\end{equation}
as a function of $S$: $\beta = \beta(S)$.
}
\\[3mm]
In (\ref{Eqn_6}), $\beta = \beta (\alpha; t, u)$. So, $\beta$ contains all the physical and dynamical information of the landslide, including, the slope, particle and fluid densities, the particle concentration and friction.
  {\it Proposition 1} asserts that for each acceleration number $S$, determined by both the mechanics and dynamics of the landslide, there exists a unique positive root $\eta$ of the dynamical drag model (\ref{Eqn_5}).

\subsection{The procedure}

The drag dynamics is the major concern here. Proper model for the (viscous or the ambient) drag coefficient is needed to automatically determine the drag during the mass flow simulations. The produce is as follows. As (\ref{Eqn_6}) defines the drag coefficient, (\ref{Eqn_5}) and (\ref{Eqn_6}) provide a mechanical method for calculating the dynamic drag in landslide. As $S$ is regulated by the physics and dynamics of the flow, $\eta$ is then independently obtained from (\ref{Eqn_5}) without any information on $\beta$. This is an advantage. However, it intrinsically contains $\beta$.  Ultimately, this $\eta$ is employed in (\ref{Eqn_6}), determining $\beta$ as a function of $S$. In principle, this addresses a grand standing problem of the drag dynamics.

\subsection{The roots of $F$}

\subsubsection{Intersecting components of $F$}

\begin{figure}[ht!]
\begin{center}
\includegraphics[width=15.cm]{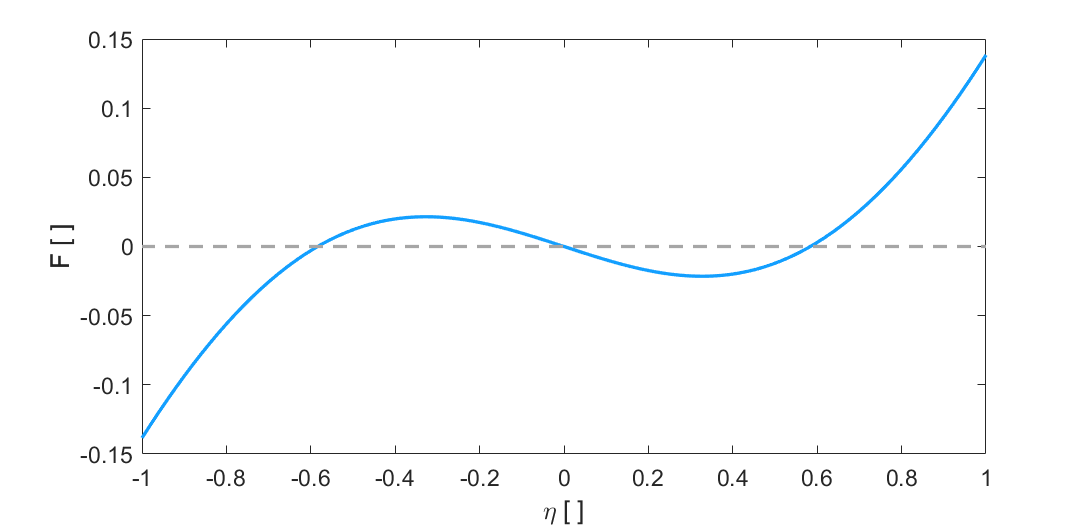}
  \end{center}
  \caption[]{Roots of the drag function $F$ as given by (\ref{Eqn_5}). There are three roots: a negative root, a zero root, and a positive root.}
  \label{Fig_1}
  \end{figure}

The real number $\eta$ that identically satisfies (\ref{Eqn_5}) is called the root of $F$.
The roots of $F$ are determined by the intersection of the linear component $F_L = S \eta$ and the transcendental component $F_T = \tanh(\eta)$ of $F$, where $F = F_L - F_T$. Since the function $F_T$ is fixed, the roots then depend on $F_L$, namely, on the values of $S$ that defines the slope of $F_L$. Thus, the roots of $F$ are entirely determined by the physics, mechanics and the dynamics of the landslide as contained in $S$. The nature of the tangent hyperbolic function suggests that as the value of $S$ increases, $F_L$ intersects $F_T$ closer to the origin. Similarly, as the value of $S$ decreases, $F_L$ intersects $F_T$ away from origin. The roots of $F$ are displayed in Fig. \ref{Fig_1} for a plausible value of $S = 0.9$. The abscissa of these intersections are the roots of $F$. There are three roots: a negative root, a zero root, and a positive root.

  \subsubsection{Root determination: analytical method}

  The model (\ref{Eqn_5}) is a composite of an algebraic and a transcendental function. So, there is no hard, straightforward rule to analytically find its roots.
This means, (\ref{Eqn_4}) cannot be solved analytically. However, we take advantage of the nature of the drag coefficient $\beta$ in constructing a meaningful and useful analytical solution. The point is, since $\beta$ is a small positive number, $\eta$ can be a small number. So, $\tanh(\eta)$ can be approximated by $\eta - \eta^3/3$. With this, (\ref{Eqn_4}) can be written as:
\begin{equation}
\displaystyle{S \eta \approx \eta - \frac{1}{3}\eta^3}.
\label{Eqn_beta1p}
\end{equation}
This says, either $\eta = 0$, implying $\beta = 0$, which effectively means that formally, zero drag is no drag. Or, we have,
\begin{equation}
\displaystyle{\eta^2 = 3(1-S)}.
\label{Eqn_beta2p}
\end{equation}
This means that the drag coefficient is fully described by the acceleration number $S$. Now, recalling the definitions of $\eta$ and $S$ from Section 3.2, for the positive net driving acceleration ($\alpha > 0$) the drag coefficient $\beta^+$ takes the form:
\begin{equation}
\displaystyle{\beta^+ = \frac{\lambda^+}{\alpha^2\,t^3}\left( \alpha\,t - u\right)},
\label{Eqn_beta3p}
\end{equation}
where, the superscript $^+$ in $\beta^+$ indicates that this drag coefficient is associated with the positive net driving acceleration, and $\lambda^+ = 3$.
This reveals that the drag is characterised by the structure $\left(\alpha\,t - u\right)$ that includes all the physical and dynamical quantities associated with the landslide motion.
\\[3mm]
{\bf Mechanism of the drag coefficient:}
\\[1mm]
In situation when $\alpha t - u = 0$, $\beta^+ = 0$. Then, (\ref{Eqn_1}) tells that, this is equivalent to disregarding the drag force. So, (\ref{Eqn_beta3p}) is mathematically fully consistent. Next, we explore what the dynamic drag coefficient (\ref{Eqn_beta3p}) tells mechanically. On the one hand, $\alpha t$ is the ideal velocity (with respect to the net driving acceleration, $\alpha$) if the landslide would move with 100\% efficiency without losing any energy in (viscous) resistance. While on the other hand, $u$ is the velocity of the landslide as the drag force consumes (some) energy in dissipation, the real situation. $\alpha t$ is the potentially maximum velocity the system would attain with the available net driving acceleration, $\alpha$; and $u$ is the effective velocity once the system experiences the energy dissipation with drag. The difference between these ideal and the real velocities, $\alpha t - u$, accounts for the net velocity loss by drag in energy dissipation as the landslide moves. So, $\alpha t - u$ is the total dissipated velocity, or the amount of speed lost to the environment. This is the measure of the induced retardation, or the inertial lag. In other words, $\alpha t - u$ is a gross measure of the landside (energy or mobility) inefficiency.  If $\alpha t - u$ increases in time, the dissipative force due to drag increases as the drag coefficient increases. This means, the lower the actual velocity, the higher the drag coefficient. Conversely, if $\alpha t - u$ decreases in time, the dissipative force due to drag is low as the drag coefficient decreases. In other words, the higher the actual velocity, the lower the drag coefficient. As revealed here, these are fascinating mechanisms inherited by the dynamical drag coefficient (\ref{Eqn_beta3p}). We will further elaborate on it while applying this drag model in simulating the landslide motion with discussion at Section 6.
\\[3mm]
The factor $\alpha^2 t^3 = (\alpha t)(\alpha t^2)$ in the denominator of (\ref{Eqn_beta3p}) closes the drag mechanism in which $\alpha t$ is the basic velocity scale of the system. So, $S_I = \left(\alpha t - u\right)/\left(\alpha t\right)$ is the relative velocity loss (or the head loss) in drag, a dimensionless quantity that characterizes the drag coefficient. We call $S_I$ the measure of energy inefficiency. The remaining part, $S_L = \alpha t^2/\lambda^+$ is the characteristic length, we call it the drag length scale. All these descriptions match very well with the fundamental definition of drag at Section 2.3.
\\[3mm]
{\bf Mechanical definition of the drag coefficient:}
\\[1mm]
With the above considerations, we can now, for the first time, formally define the mechanical drag coefficient as the ratio between the landslide inefficiency $(S_I)$ and the drag length $(S_L)$:  $\displaystyle{\beta_S = \frac{S_I}{S_L}}$. This appears to be a natural definition of the drag coefficient.  In short, the drag is the measure of (immobility, or ) inefficiency (per unit length).

\subsection{The dynamical drag equation for negative net driving acceleration: $\alpha < 0$}

For $\alpha < 0$, the situation becomes complex, because, then, we need to use the information on the time and velocity as the flow changes its regime from the positive net driving acceleration ($\alpha > 0$) to the negative net driving acceleration ($\alpha < 0$), say at $\left(t_0, u_0\right)$. Then, at this instance, the behaviour changes completely.
For this, the exact-analytical solution of (\ref{Eqn_1}) is given by (Pudasaini and Krautblatter, 2022):
\begin{equation}
\displaystyle{u = -\sqrt{\frac{\alpha^{-}}{\beta}}\tan\left [\left(\sqrt{\alpha^{-}} (t - t_0)\right)\sqrt{\beta} - \tan^{-1}\left({u_0}\sqrt{\frac{\beta}{\alpha^{-}}}\right)\right ]},
\label{Eqn_beta1m}
\end{equation}
with the sign of $\alpha$ switched to negative, $\alpha^- = - \alpha, \alpha^- > 0$, which will be employed to construct the dynamical drag coefficient for this situation.
This can be re-cast in equivalent form as:
\begin{equation}
\displaystyle{\left(\frac{u}{\sqrt{\alpha^-}}\right) \sqrt{\beta} = -\tan\left [\left(\sqrt{\alpha^{-}} (t - t_0)\right)\sqrt{\beta} - \tan^{-1}\left(\frac{u_0}{\sqrt{\alpha^{-}}}\sqrt{\beta}\right)\right ]},
\label{Eqn_beta2m}
\end{equation}
which is more complicated than (\ref{Eqn_2}).
Now, with the definition:
\begin{equation}
\displaystyle{T = t-t_0,\,\, S = \frac{u/T}{\alpha^-},\,\, S_0 = \frac{u_0/T}{\alpha^-},\,\, \eta = \left(\sqrt{\alpha^-}\,T\right)\sqrt{\beta}},
\label{Eqn_beta3m}
\end{equation}
(\ref{Eqn_beta2m}) can be written in a compact form as:
\begin{equation}
\displaystyle{S \eta = - \tan \left[\eta - \tan^{-1}\left(S_0 \eta\right)\right]}.
\label{Eqn_beta4m}
\end{equation}
As before, for the situation of the negative net driving acceleration, $S$ is called the acceleration number, and $S_0$ is called the terminal acceleration number, because, it acts at the instance of the regime change from the positive net driving acceleration to the negative net driving acceleration at $\left(t_0, u_0\right)$. With the trigonometric identity $\tan(-\theta) = - \tan(\theta)$, (\ref{Eqn_beta4m}) takes the form:
\begin{equation}
\displaystyle{S \eta = \tan \left[-\eta + \tan^{-1}\left(S_0 \eta\right)\right]}.
\label{Eqn_beta5m}
\end{equation}
Applying the arctangent on both sides, re-arranging the resulting expression, and simplifying with the sum of the arctangents yields:
\begin{equation}
\displaystyle{\eta = -\left[ \tan^{-1}\left(S\eta\right) - \tan^{-1}\left(S_0\eta\right)\right] = -\tan^{-1}\left[\frac{S\eta - S_0\eta}{ 1 + S\eta S_0\eta} \right]}.
\label{Eqn_beta6m}
\end{equation}
For small $\eta$, this can further be approximated and expressed in a series form, resulting in:
\begin{equation}
\displaystyle{\eta \approx -\tan^{-1}\left[{\left(S - S_0\right)\eta} \right] \approx
-\left[ \left(S - S_0\right)\eta -\frac{1}{3}\left(S - S_0\right)^3\eta^3 \right] }.
\label{Eqn_beta7m}
\end{equation}
This says, either $\eta = 0$, implying $\beta = 0$, which effectively means that formally, zero drag is no drag. Or, we have,
\begin{equation}
\displaystyle{\eta^2 = 3\frac{\left(1 + S - S_0\right)}{\left(S - S_0\right)^3}}.
\label{Eqn_beta8m}
\end{equation}
This shows that the drag coefficient is fully described by the acceleration numbers. So, the acceleration number controls the drag dynamics.
Recalling the definitions from (\ref{Eqn_beta3m}) and $\alpha^- = -\alpha$, for the negative net driving acceleration ($\alpha < 0$), the dynamical drag coefficient $\beta^-$ is given by:
\begin{equation}
\displaystyle{\beta^- = \frac{-\lambda^-\alpha}{\left(u - u_0\right)^3} \left[\left(u - u_0\right) - \alpha \left(t - t_0\right)\right]},
\label{Eqn_beta9m}
\end{equation}
where, the superscript $^-$ in $\beta^-$ designates that this drag coefficient is associated with the negative net driving acceleration $(\alpha < 0)$, and $\lambda^- = 3$. This indicates that the drag is characterized by the structure of the nature $\left(\alpha\,T - U\right)$, where $T = t - t_0, U = u - u_0 $, that includes all the physical and dynamical quantities associated with the landslide motion. Then, (\ref{Eqn_beta9m}) can be written in a compact form as:
\begin{equation}
\displaystyle{\beta^- = \frac{\lambda^-\alpha}{U^3} \left(\alpha\,T - U\right)}.
\label{Eqn_beta9ma}
\end{equation}
For the order of magnitude estimate, $U$ at the denominator can be approximated by $\alpha\,T$, yielding:
\begin{equation}
\displaystyle{\beta^- = \frac{\lambda^-}{\alpha^2 T_c^3}\left(\alpha\,T - U\right)},
\label{Eqn_beta9mb}
\end{equation}
where, $T_c = T + C$, and $C$ is the compensator of this approximation which may be close to $t_0$.
The mechanism of the drag coefficient $\beta^{+}$ explained at Section 3.4.2 also applies to $\beta^{-}$.

 \subsection{The unified evolutionary drag coefficient}

 From (\ref{Eqn_beta3p}) and (\ref{Eqn_beta9mb}), the unified, evolutionary, mechanical drag coefficient $\beta$ is given by:
\begin{equation}
\displaystyle{
\beta =
\begin{cases}
\displaystyle{\beta^{+} = \frac{\lambda^+}{\alpha^2\,t^3}\left( \alpha\,t - u\right)}, & \text{if}\,\,\, \alpha > 0, \\[5mm]
\displaystyle{\beta^{-} = \frac{\lambda^-}{\alpha^2 T_c^3} \left(\alpha\,T - U\right)}, & \text{if}\,\,\, \alpha < 0,
\end{cases}
}
\label{Eqn_beta10-1}
\end{equation}
where, $\alpha = g\sin(\zeta) - (1-\gamma)\alpha_s g \cos(\zeta)\mu$ is the net driving acceleration with the parameter definitions at Section~3.1, and $T_c = t - t_0 + C, U = u - u_0$. It is phenomenal that both branches of $\beta$ take the same universal form demonstrating the mathematical-physical consistency of the derived drag model.  By definition, $\beta$ is a positive quantity.
 So, in application its magnitude can be considered.
Although from simple derivation $\lambda^{+} = 3$, and $\lambda^{-} = 3$, other values of $\lambda^{\pm}$ in the vicinity of 3 may capture higher order effects.
\\[3mm]
Two ideas played important role in constructing the elegant analytical drag coefficients in (\ref{Eqn_beta10-1}). First, the definition and the structure of $S$ and the auxiliary variable $\eta$. Second, the realization that the dynamical drag coefficients are small positive numbers.
\\[3mm]
The drag model (\ref{Eqn_beta10-1}) can be applied for both the deformable body and non-deformable object.
It is less likely in real situation, however, if $\alpha$ approaches zero from either side, then, in practice, one can use the proper regularization by considering the $\beta^+$ and $\beta^-$ branches for $\alpha$ sufficiently away from zero.
However, note that, depending on how we approximate the arctangent in (\ref{Eqn_beta6m}), and to each components or the sum, we can obtain different expressions for $\eta^2$ in (\ref{Eqn_beta8m}) resulting in different $\beta^-$ in (\ref{Eqn_beta9m}). Numerical root finding methods could also be applied to (\ref{Eqn_5}) and (\ref{Eqn_beta5m}).
\\[3mm]
{\bf Explicit the viscous effect:}
 For rapid flows and bluff-type objects, the skin-drag can be independent of viscosity as the inertial force dominates the viscous force. However, for relatively moderate flows with streamlined-objects, the skin drag may be weakly related to the viscosity, typically proportional to the one-fifth-power of the dynamic viscosity  (Hoerner, 1965; Roskam, 1985; Flack, 2018; Fan, 2019; Anderson, 2023). Landslide moving through a viscous fluid may be considered as a streamlined-object.
\\[3mm]
However, to explicit the role of viscosity of the ambient fluid {$\eta_f$}, for practical purpose, we extract the viscous effect from the total drag coefficient. Without loss of generality, the drag coefficient $\beta$ can be split into the pressure (form)- and the skin- drag coefficients. A careful formal derivation suggests that a factor $\left (1 + \xi\eta_f^\chi\right )$, where, $\chi \sim 1/5$, emerges in (\ref{Eqn_beta10-1}).  Therein, $\xi$ includes the geometrical and the dynamical constraints, namely, the ratio between skin-friction coefficient $C_f$ and the geometric constant (base-line drag coefficient) $C_0$; the ratio between the reference wet area $S$ to the reference projected area $A$, and the inertial contribution $1/(\rho U L)^{1/5}$ with some typical free-stream velocity $(U)$ and the length-scale $(L)$: $\displaystyle{\xi = \frac{C_f}{C_0}\frac{S}{A}\frac{1}{(\rho U L)^{1/5}}}$ (Hoerner, 1965; Roskam, 1985; Flack, 2018; Fan, 2019; Anderson, 2023). So, depending on the flow configuration, $\xi$ could take values on the order of 1. Moreover, $\chi$ characterizes the viscous strength that may deviate away from 1/5. That, however, depends on the flow (fluid) and the geometry of the landslide, namely, the flow configuration, and whether the flow is turbulent-type or laminar-type. Again, without loss of generality, for ease, $\xi$ could be embedded in to $\eta_f$ and adjust its value, but leave $\xi$ therein just to reflect that $\xi$ has the dimension of that of $1/\eta_f^\chi$, thus, making $\xi \eta_f^\chi$ a dimensionless quantity. This makes $\left (1 + \xi\eta_f^\chi\right) \beta$ the viscosity-consistent drag coefficient. In this formulation, $C_0$ is implanted in $\beta$.
\\[3mm]
With these considerations, from (\ref{Eqn_beta10-1}), yields the evolutionary drag coefficient with explicit viscous effects:
\begin{equation}
\displaystyle{
\beta =
\begin{cases}
\displaystyle{\beta^{+} = \frac{\lambda^+}{\alpha^2\,t^3}\left (1 + \xi\eta_f^\chi\right)\left( \alpha\,t - u\right)}, & \text{if}\,\,\, \alpha > 0, \\[5mm]
\displaystyle{\beta^{-} = \frac{\lambda^-}{\alpha^2 T_c^3}\left (1 + \xi\eta_f^\chi\right) \left(\alpha\,T - U\right)}, & \text{if}\,\,\, \alpha < 0.
\end{cases}
}
\label{Eqn_beta10}
\end{equation}
Usually, the main dense core of (the subaerial and the submarine) landslides and avalanches are enveloped (covered) by a layer of turbulent dense cloud of suspended fine particles mixed with the surrounding fluid. These ambient fluids are collision-dominant, suspended clouds of turbulent, colloidal, fine particles. The viscosity of such fluids depends on the particle concentration than can be low (Einstein, 1906), semi-dilute (Thomas, 1965; Batchelor, 1977) or highly crowded (Krieger and Dougherty, 1959). As such, these fluids may have relatively high particle density, conceivably, resulting in the high particle-laden fluid viscosity, possibly on the order of 0.01 Pas, such as that of the powder cloud and turbidity currents, or even higher. So, in practical applications, the order of magnitude estimates of $\xi$ and $\eta_f$ influence the strength of the viscous effect of the ambient fluid in determining the drag coefficient.

\section{The dynamical model equations with evolutionary, mechanical drag}

\subsection{The dynamical model equations}

In real simulations, the drag values will be automatically determined by the flow dynamics (through the balance equations) and the associated mechanisms.
With the unified evolutionary drag coefficient $\beta$ as described by (\ref{Eqn_beta10}), the mass and momentum balance equations (Pudasaini and Mergili, 2025) for the granular landslide motion down a two-dimensional channel are given, respectively, by:
\begin{equation}
\displaystyle{\frac{\partial h}{\partial t} + \frac{\partial \left(hu \right)}{\partial x} = 0},
\label{Eqn_mass}
\end{equation}
\begin{equation}
\displaystyle{\frac{\partial (hu)}{\partial t} + \frac{\partial}{\partial x}\left[hu^2 + K g^z\frac{h^2}{2} \right] = h \left [g^x -\frac{u}{|u|}g^z \tan(\delta)
- g^z\frac{\partial b}{\partial x}
-\beta\,u|u|\right]},
\label{Eqn_moment}
\end{equation}
where $g^x, g^z$ are the components of gravity acceleration along ($x$-direction) and perpendicular to the slope, $b = b(t, x)$ is the basal topography,
and $K = K (\phi,\delta)$ is the earth pressure coefficient as a function of the internal ($\phi$) and basal ($\delta$) friction angles (Pudasaini and Hutter, 2007).
 The model equations (\ref{Eqn_mass})-(\ref{Eqn_moment}) can be directly extended to geometrically three-dimensional flows (Pudasaini et al., 2024).

\subsection{The central question}

What really is the functionality of drag coefficient, and how do we know which drag coefficient represents the physical reality? Moreover, why should the drag coefficient be a constant for the entire flow for a dynamically evolving landslide? This is a great open question. Overwhelming attempts have been made to interpret and justify the simulation results with some conveniently chosen constant drag coefficients (Brufau et al., 2000; Chen et al., 2006; Hungr and McDougall, 2009; Christen et al., 2010; Frank et al., 2015; Melo et al., 2018; Shugar et al., 2021; Meyrat et al., 2023; Mergili et al., 2025; Sattar et al., 2025) such that the observed laboratory and the field data are simulated satisfactorily. However, this is only a parameter calibration with some kind of satisfaction rather than representing any real physics and the mechanics of the mass flow. The simulation results presented below address these questions.

\subsection{Numerical method, simulation set-up and initial conditions}

To explore the effect of the analytical drag and its control over the motion, spreading, run-out and the deposition of landslides or avalanches, we consider granular mass flows with constant drag coefficients used exclusively in literature, and the new drag model developed here. Simulations are performed by solving the dynamical model equations (\ref{Eqn_mass})-(\ref{Eqn_moment}) for geometrically two-dimensional flows down a channel with the high resolution and very efficient TVD-NOC numerical method (Tai et al., 2002; Pudasaini and Hutter, 2007), which have been extensively applied in mass flow simulations (Pudasaini and Mergili, 2019; Shugar et al., 2021; Mergili et al., 2025; Sattar et al., 2025; Feng et al., 2026). Here, we consider an inclined channel with slope angle $\zeta = 45^\circ$ until $x = 80$ m and $\zeta = 0$ for $x > 80$ m, defining the transition, run-out and deposition region (Fig. \ref{Fig_3}). The internal and the basal friction angles are $\phi = 33^\circ$ and $\delta = 27^\circ$. The initial triangular mass of the frontal depth of 5 m lies between 10 m and 30 m in the upper part of the slope. This mass is released with the initial velocity zero. Similar set-ups can be found in Pudasaini and Hutter (2007).
{For the simulation results presented below at Section 5, for simplicity, the factor $\left (1 + \xi\eta^\chi_f\right)$ is included in $\lambda$ in (\ref{Eqn_beta10}).}

\section{Simulation results}

\subsection{Landslide dynamics with globally constant drag coefficient}

\begin{figure}[t!]
\begin{center}
\includegraphics[width=15.cm]{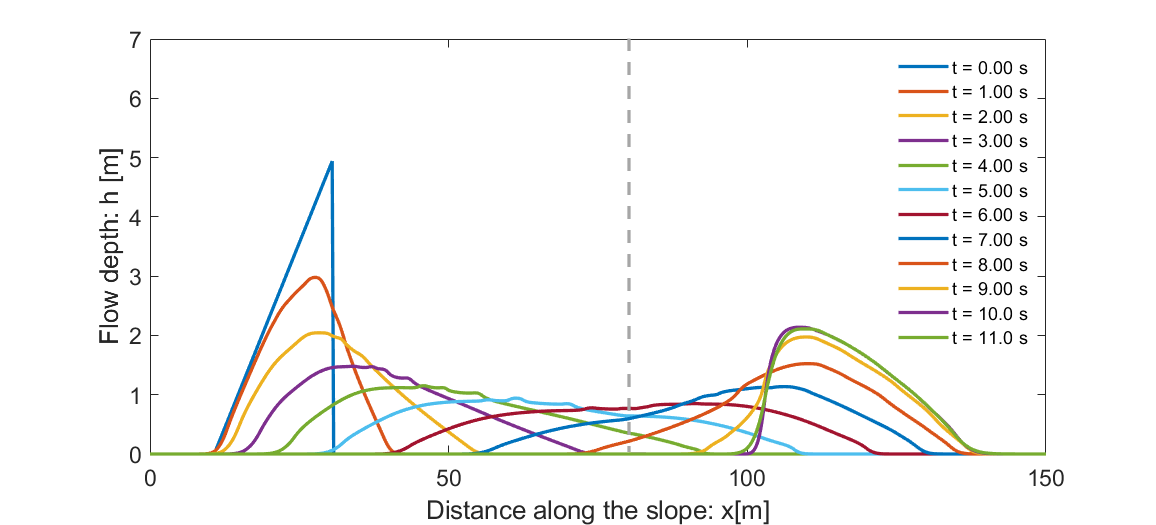}
  \end{center}
  \caption[]{Landslide dynamics with constant drag coefficient $\beta = 0.0002$. The gray dashed line at $x = 80$ m indicates the transition to the run-out (change in slope from $\zeta = 45^\circ$ to $\zeta = 0^\circ$).}
  \label{Fig_3}
\begin{center}
\includegraphics[width=15.cm]{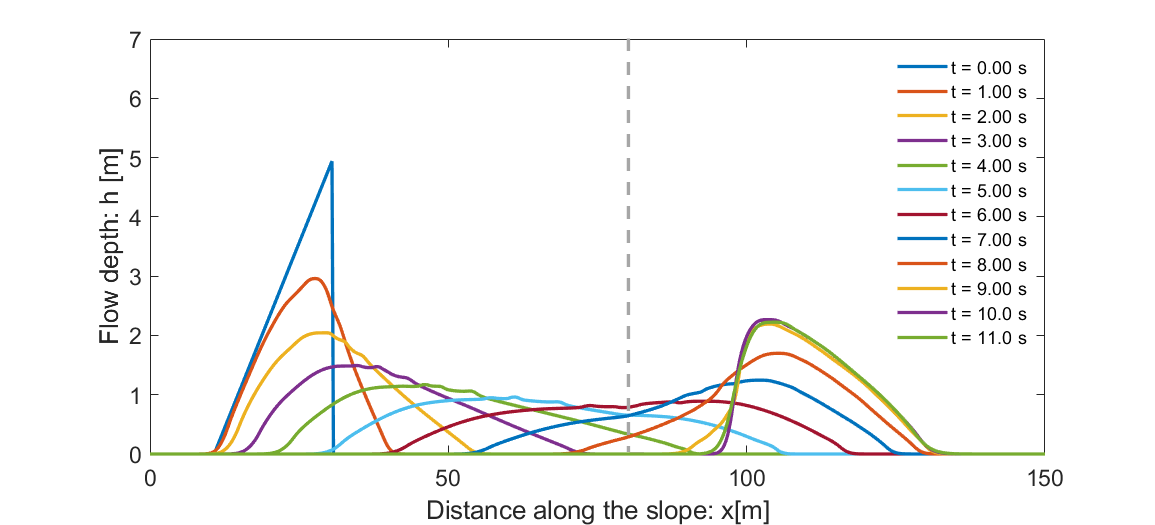}
  \end{center}
  \caption[]{Landslide dynamics with constant drag coefficient $\beta = 0.002$. The gray dashed line at $x = 80$ m indicates the transition to the run-out (change in slope from $\zeta = 45^\circ$ to $\zeta = 0^\circ$).}
  \label{Fig_4}
\end{figure}

\begin{figure}[t!]
\begin{center}
\includegraphics[width=15.cm]{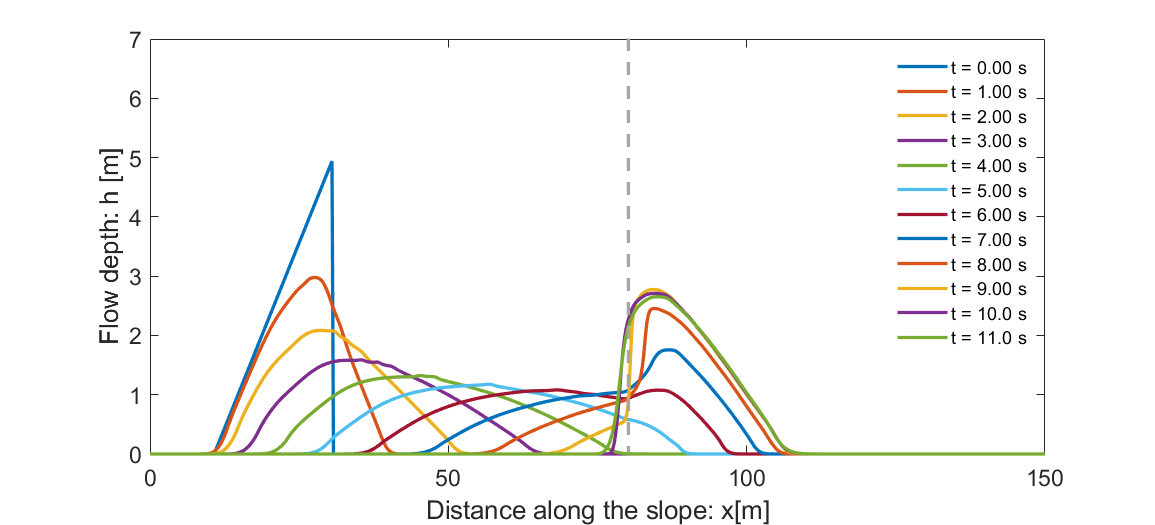}
  \end{center}
  \caption[]{Landslide dynamics with constant drag coefficient $\beta = 0.02$. The gray dashed line at $x = 80$ m indicates the transition to the run-out (change in slope from $\zeta = 45^\circ$ to $\zeta = 0^\circ$).}
  \label{Fig_5}
\begin{center}
\includegraphics[width=15.cm]{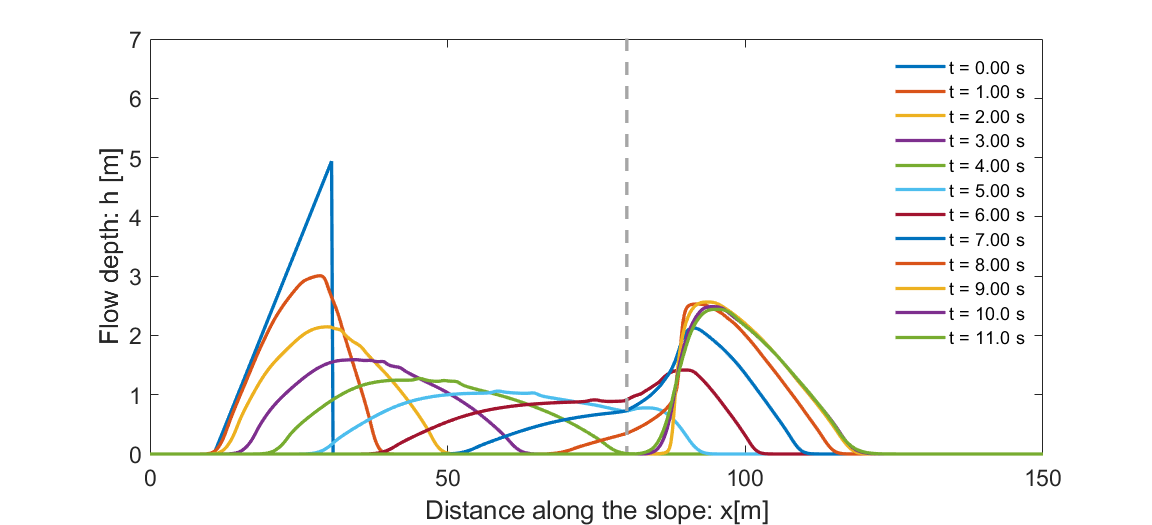}
  \end{center}
  \caption[]{Landslide dynamics with evolutionary analytical drag coefficient $\beta$ as given by (\ref{Eqn_beta10}). The gray dashed line at $x = 80$ m indicates the transition to the run-out (change in slope from $\zeta = 45^\circ$ to $\zeta = 0^\circ$).}
  \label{Fig_6}
\end{figure}

The two types of simulation results, with constant drag coefficients and the dynamically evolving analytical drag coefficient, appear to be fundamentally different: the way the mass propagates down the slope, its stretching, transition to the run-out and the deposition morphology.
\\[3mm]
Figure \ref{Fig_3}, Fig. \ref{Fig_4}, Fig. \ref{Fig_5} display the landslide dynamics and deposition for globally constant drag coefficients $\beta = 0.0002, 0.002, 0.02$, respectively, representing almost no drag, mild drag and the strong drag. These or similar values are exclusively employed in simulations (Kattel et al., 2016; Pudasaini and Fischer, 2020a; Pudasaini and Krautblatter, 2022; Mergili et al., 2025). With the constant drag, the propagating masses on the inclined part show very tapering fronts, all the way down to the transition, also in the run-out plane. Even with the mild constant drag (Fig. \ref{Fig_4}), the travel reach is substantially longer. However, with the strong constant drag (Fig. \ref{Fig_5}), two further aspects draw our attention. First, as soon as the mass reaches the transition, its deceleration is strongly rapid, like a sudden halt, forcefully stopping the front, probably much earlier than one would anticipate. Second, the successive rear geometry of the depositing mass is almost perpendicular to the sliding surface. These geometrical forms are less realistic as the laboratory experiments of mass flows show diffused shock structures rather than such a perfect bore (Pudasaini and Hutter, 2007).
\begin{figure}[t!]
\begin{center}
\includegraphics[width=15.cm]{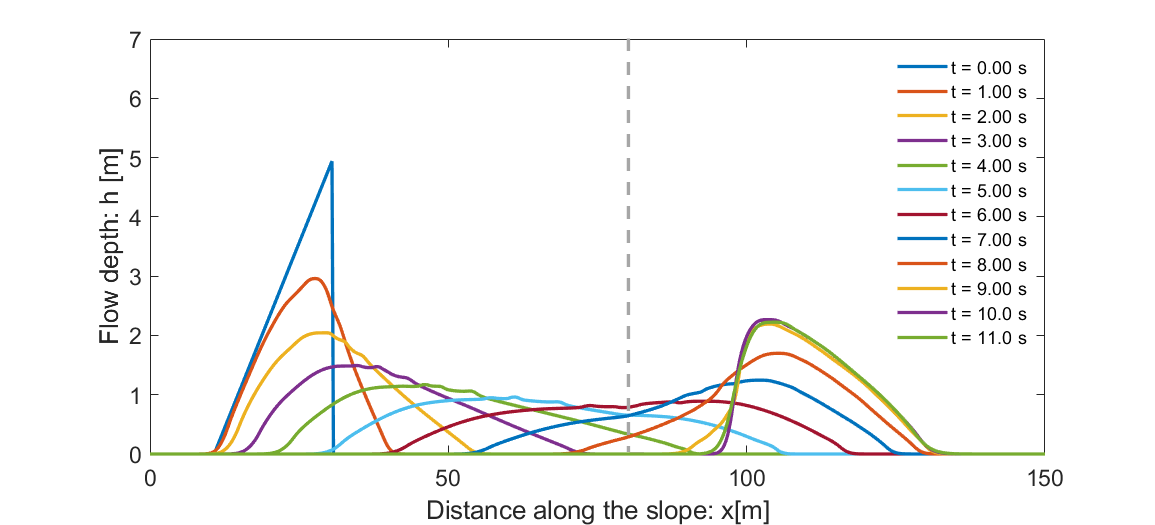}
\includegraphics[width=15.cm]{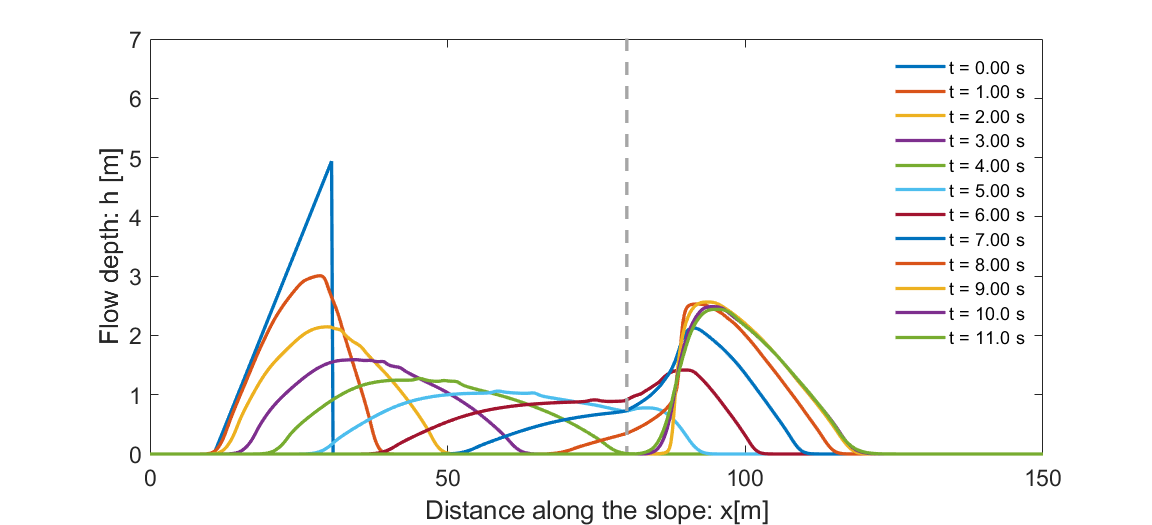}
  \end{center}
  \caption[]{Landslide dynamics with: (A) constant drag coefficient $\beta = 0.002$, and (B) evolutionary analytical drag $\beta$ as given by (\ref{Eqn_beta10}). The gray dashed line at $x = 80$ m indicates the transition to the run-out (change in slope from $\zeta = 45^\circ$ to $\zeta = 0^\circ$).}
  \label{Fig_7}
\begin{picture}(0,0)
\put(120,435){{\bf A}}
\put(120,240){{\bf B}}
\end{picture}
\end{figure}

\subsection{Landslide dynamics with mechanically evolving drag coefficient}

Figure \ref{Fig_6} shows the landslide motion with the new, dynamic, analytical drag coefficient model (\ref{Eqn_beta10}) with $\left(t_0, u_0\right)~\!=~\!(4, 10)$ as estimated from the frontal dynamics at the transition.
The results in Fig. \ref{Fig_6} appear to be much more realistic. There are couple of reasons for that. First, the drag coefficient is now explained physically. Second, we know that the flow dynamics changes during the mass propagation. The landslide can accelerate, may develop into a steady state, and then decelerate, and finally the motion ceases. So, from this perspective, the drag coefficient must also evolve in harmony with such a complex flow dynamics. The globally constant drag coefficient, from the physical stand, appears to be meaningless. Third, in Fig. \ref{Fig_6}, the fronts are properly controlled, rather than very tapered. This is a behaviour of the frictional granular mass flows (Pudasaini and Mergili, 2025). This continues all the way to the transition. As the mass transits to the run-out zone, the front still propagates gently. In the rear, diffused shock develops as often observed deposition structures in granular mass flows (Pudasaini and Hutter, 2007). Figure \ref{Fig_7} directly compares the landslide dynamics with constant drag coefficient $\beta = 0.002$ and with evolutionary analytical drag $\beta$ as given by (\ref{Eqn_beta10}). The differences are evident and large, in the overall dynamics and deposition patterns.

\section{Discussion}

Now, we present discussion on the dynamic evolution of the mechanical drag (\ref{Eqn_beta10}), its strength and scope in mass flow simulation.

\subsection{Dynamic evolution of the mechanical drag}

\begin{figure}[!htbp]
\begin{center}
\includegraphics[width=6.4cm, height=5cm]{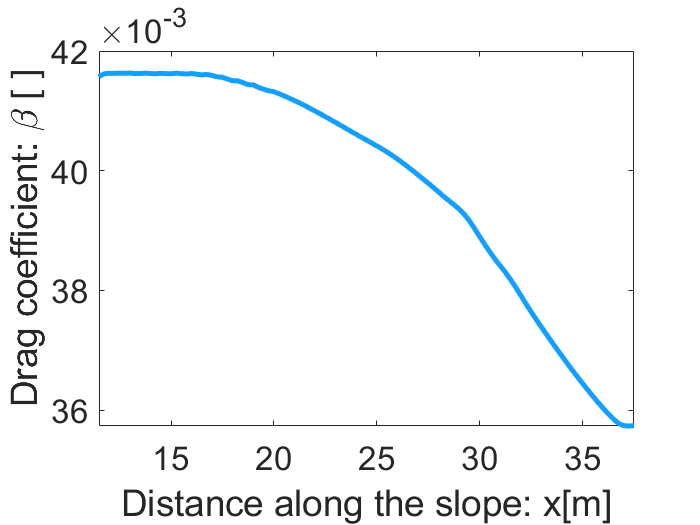}
\includegraphics[width=6.4cm, height=5cm]{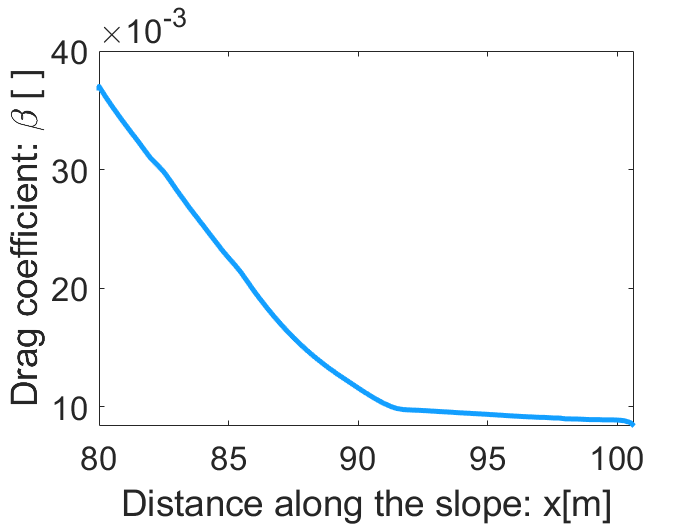}\\[-4mm]
\includegraphics[width=6.7cm, height=5cm]{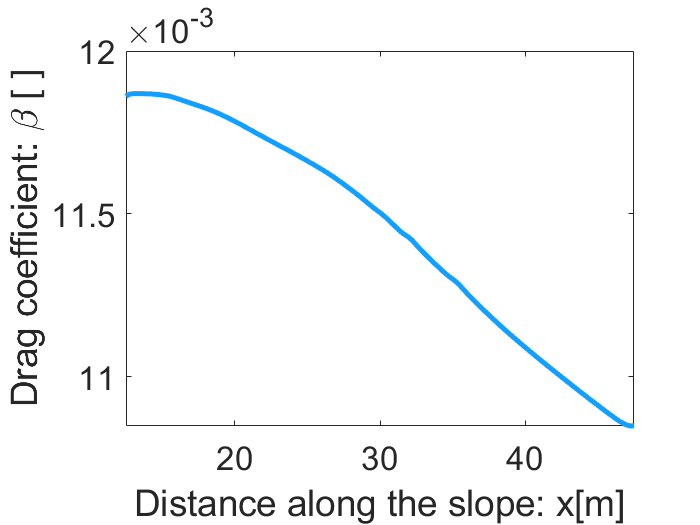}
\includegraphics[width=6.4cm, height=5cm]{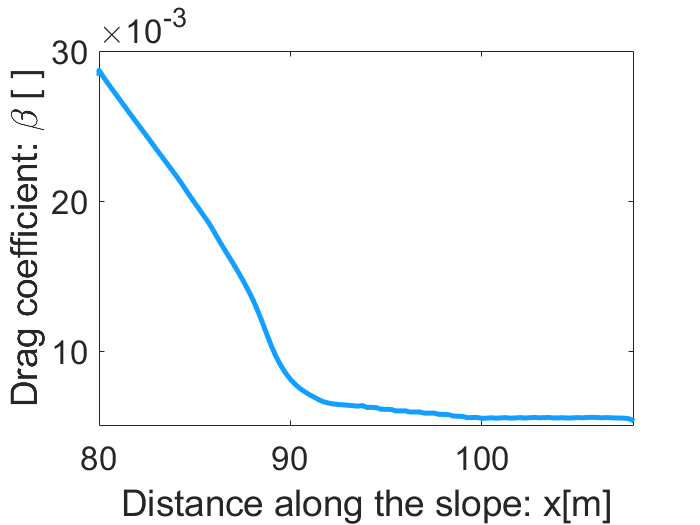}\\[-4mm]
\includegraphics[width=6.5cm, height=5cm]{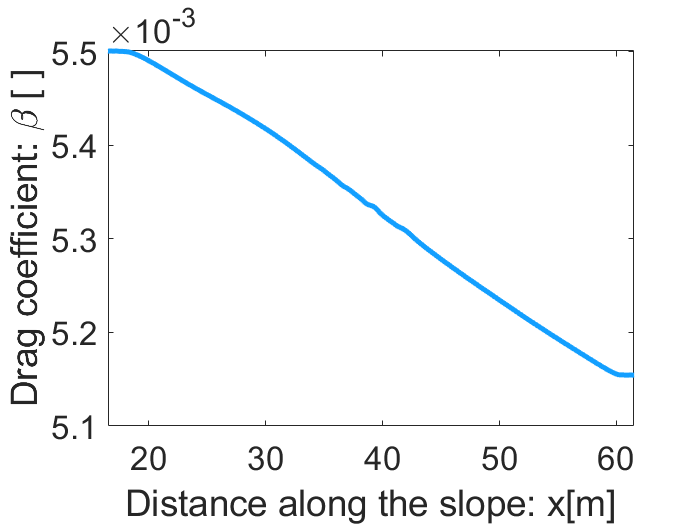}
\includegraphics[width=6.4cm, height=5cm]{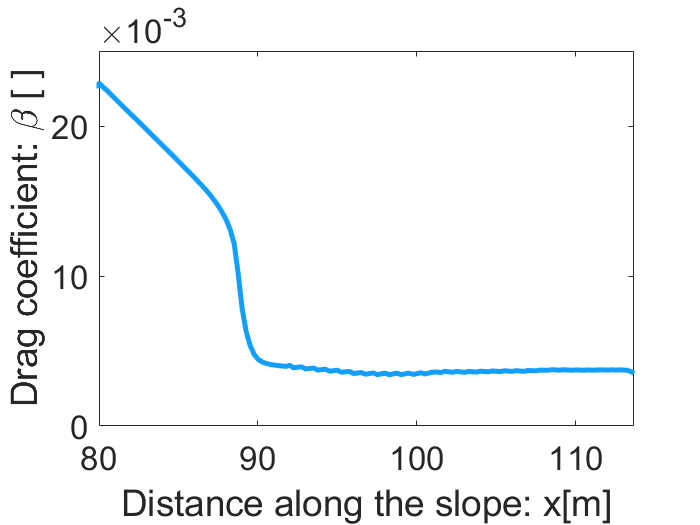}\\[-4mm]
\includegraphics[width=6.7cm, height=5cm]{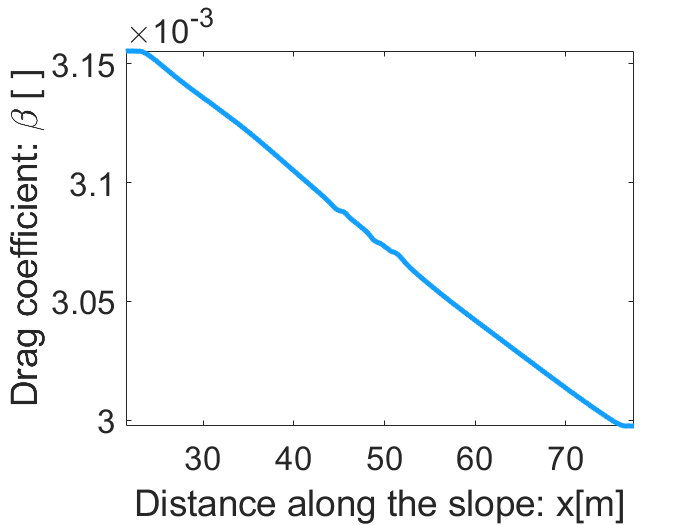}
\includegraphics[width=6.4cm, height=5cm]{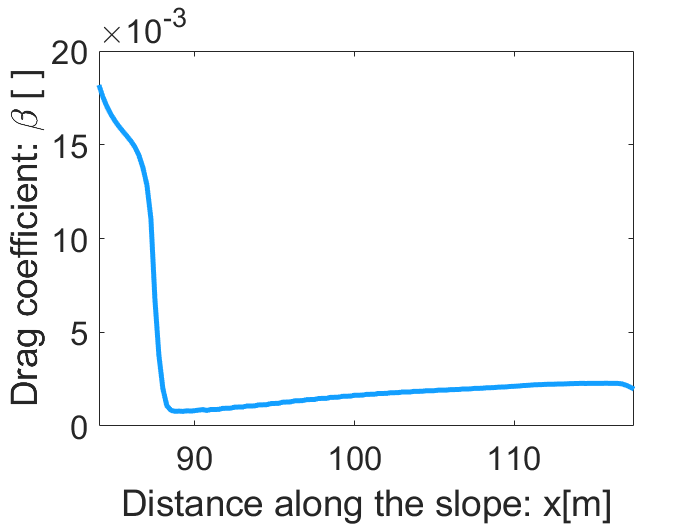}\\[-4mm]
\hspace{3mm}
\includegraphics[width=6.4cm, height=5cm]{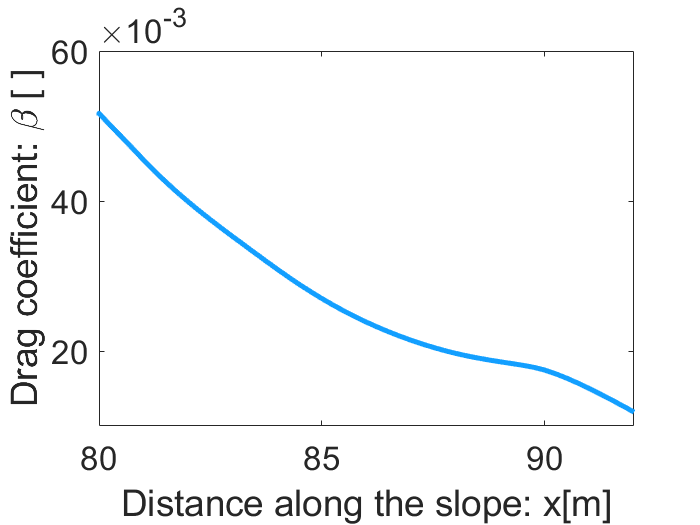}
\includegraphics[width=6.4cm, height=5cm]{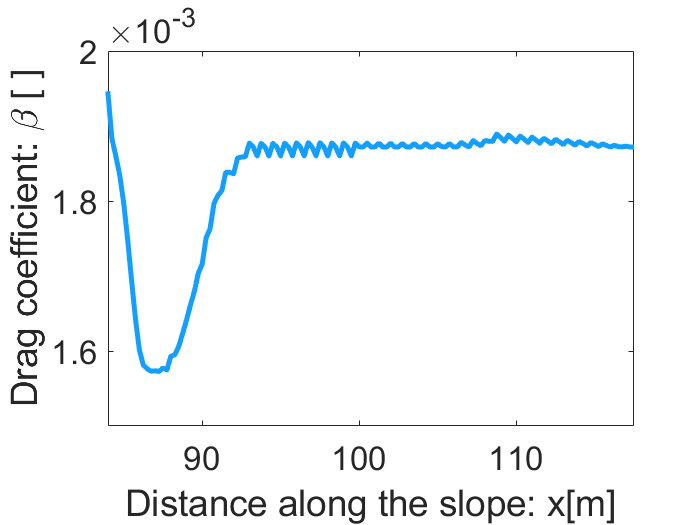}
\begin{picture}(0,0)
\put(-265,636){{\bf A:} $t = 1$ s}
\put(-265,505){{\bf B:} $t = 2$ s}
\put(-265,374){{\bf C:} $t = 3$ s}
\put(-265,243){{\bf D:} $t = 4$ s}
\put(-265,112){{\bf E:} $t = 5$ s}
\put(-80, 636){{\bf F:} $t = 6$ s}
\put(-80, 505){{\bf G:} $t = 7$ s}
\put(-80, 374){{\bf H:} $t = 8$ s}
\put(-80, 243){{\bf I:} $t = 9$ s}
\put(-80, 112){{\bf J:} $t = 10$ s}
\end{picture}
  \end{center}
  \vspace{-6mm}
  \caption[]{Time evolution of the analytical drag $\beta$ as given by (\ref{Eqn_beta10}) associated with the profiles in Fig. \ref{Fig_6}. The distance in each panel indicates the current landslide location of interest with effective drag revealing extraordinarily non-linear features with automatically evolving dynamics.}
  \label{Fig_8}
\end{figure}
\begin{figure}[!htbp]
\begin{center}
\includegraphics[width=6.4cm, height=5cm]{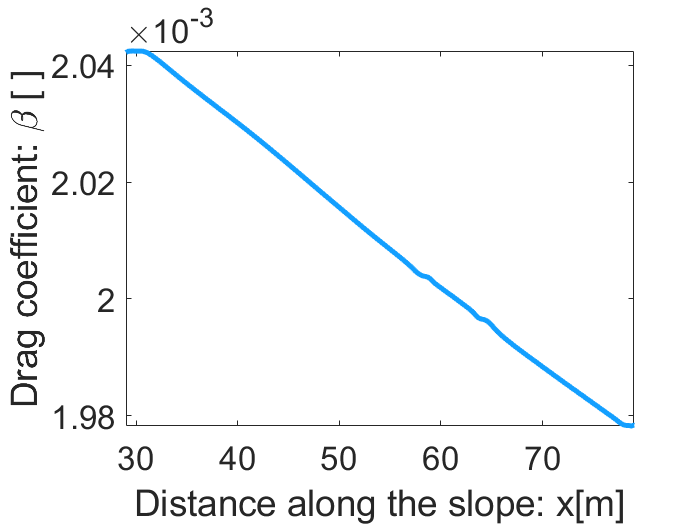}
\includegraphics[width=6.4cm, height=5cm]{MechDrag-PaperBeta-t5.png}\\[-4mm]
\includegraphics[width=6.5cm, height=5cm]{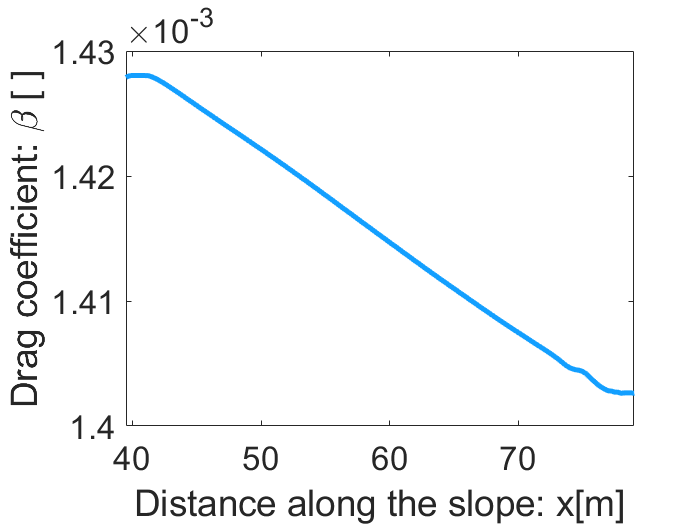}
\includegraphics[width=6.4cm, height=5cm]{MechDrag-PaperBeta-t6.png}\\[-4mm]
\includegraphics[width=6.6cm, height=5cm]{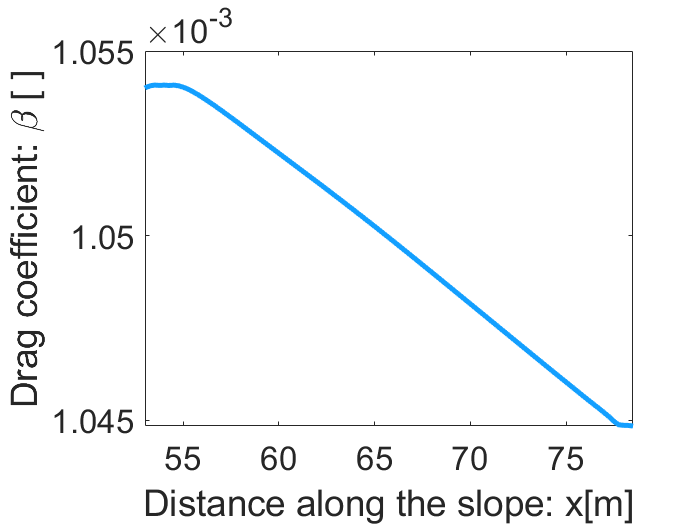}
\includegraphics[width=6.4cm, height=5cm]{MechDrag-PaperBeta-t7.png}\\[-4mm]
\includegraphics[width=6.8cm, height=5cm]{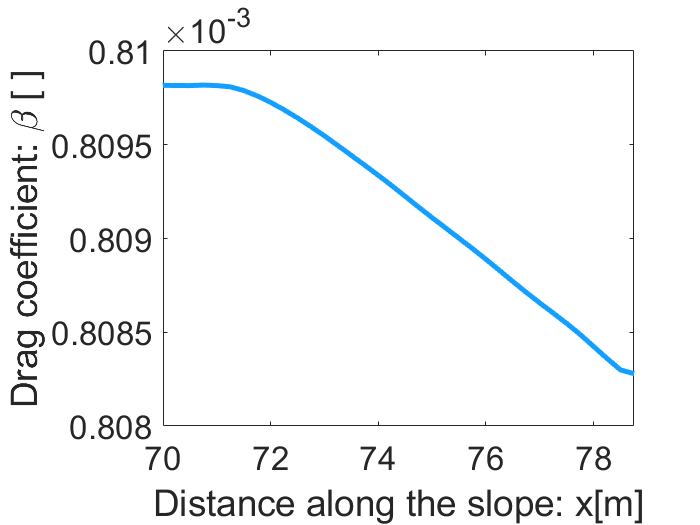}
\includegraphics[width=6.4cm, height=5cm]{MechDrag-PaperBeta-t8.png}
\begin{picture}(0,0)
\put(-265,505){{\bf E:} $t = 5$ s}
\put(-265,374){{\bf F:} $t = 6$ s}
\put(-265,243){{\bf G:} $t = 7$ s}
\put(-265,112){{\bf H:} $t = 8$ s}
\end{picture}
  \end{center}
  \vspace{-6mm}
  \caption[]{Time evolution of the analytical drag $\beta$ as given by (\ref{Eqn_beta10}) associated with the profiles in Fig. \ref{Fig_6}. The left and right panels are for the masses lying on the inclined and the run-out sectors of the slope, respectively, as indicated by panels E$-$H corresponding to those in Fig. \ref{Fig_8}. The jump in the drag coefficients at the transition are large.}
  \label{Fig_9}
\end{figure}
Figure \ref{Fig_8} displays the time evolution of the analytical drag coefficient $\beta$ for $t = 1 - 10$ s associated with the profiles in Fig. \ref{Fig_6}. The results for $t = 11, 12$ s remain virtually the same. The distance in each panel indicates the current location of the landslide with effective drag, the region of interest with highest relevance.
\\[3mm]
The comprehensive dynamics of the analytical drag in Fig. \ref{Fig_8} is very interesting and impressive. At $t = 1$ s, the rear and front are at positions 11 m and 37 m where the drag is effective. As the front moves faster than the rear and the speed of the main body lies in between, the drag coefficient is the lowest at the front and monotonically increases towards the rear. Such a dynamics persists consistently until $t = 4$ s, because, until this time, the mass lies entirely in the inclined channel where the net driving acceleration is positive. However, as the mass accelerates downslope, from $t = 2$ s to $t = 3$ s to  $t = 4$ s, with their marching rears and fronts (at (12, 47) m, (16, 61) m, (21, 78) m, respectively), the detailed dynamics is evolving differently, and the overall magnitudes are also decreased strongly, by more than an order of magnitude.
\\[3mm]
The situation changes spectacularly as the mass begins to transit into the run-out region at $t = 5$ s until $t = 8$~s. First, we analyze the status for $t = 5$ s with the rear and the front at (28, 92) m. At this instance, the landslide body lies partly in the inclined, (28, 79) m; and partly in the run-out zone, (80, 92) m. In the run-out zone, the mass is decelerating with the negative net driving acceleration. So, the drag coefficient increased heavily at transition, then, immediately dropped strongly towards the front head as the mass propagates forward. There is a huge contrast between this and the previous panels ($t = 1, 2, 3, 4$ s).
For the part of the landslide body that lies in the inclined channel ($x < 80$ m) the drag coefficients are shown in Fig. \ref{Fig_9}.
\\[3mm]
In Fig. \ref{Fig_8}E a completely new situation re-emerges than that for $t = 4$ s in Fig. \ref{Fig_8}D with sudden rise of more than an order of magnitude in the drag coefficient.
Following the regime change, here begins the second episode of the drag game.
This is counter intuitive. This occurred due to abrupt deceleration, however, its intensity decreases in the downstream and in the later time. Yet, these special behaviours are fully compatible with the model (\ref{Eqn_beta10}).
Our analysis shows that the system must re-set here with the regime change information provided by the previous sector, namely, the regime change time and velocity $(t_0, u_0)$. For $\alpha > 0$, the re-set is defined by $T = t-t_0$ and $U = u-u_0$ as the subsequent dynamics is controlled by the evolving velocity $u$. As $u$ evolves differently in the decelerating regime than in the accelerating regime, $\beta^-$ evolves completely differently than $\beta^+$. From the contemporary perspective, this is unusual, and is a strikingly novel phenomenon.
\\[3mm]
At $t = 6$ s, the mass lies in (38, 101) m. At this instance, the interesting dynamics lies in (80, 101) m. In the sector (80, 91) m, the drag coefficient drops heavily, because, after the transition to the run-out, the front is still moving. For (91, 101) m, the drag coefficient decreases slowly as that part of the mass is still moving with a moderate speed.
 However, the main change occurred at $x = 91$ m. Such strong local changes resulted because of the rapidly evolving drag coefficient as ruled by the flow transition and changes in the physical parameters.
 \\[3mm]
The evolutionary dynamics continues for $t = 7, 8, 9$ s with rears and fronts at (52, 109) m, (68, 115) m, (86, 119) m, respectively, however, with significantly decreased intensities and elevated non-linearities of the drag coefficient. This is, because, the deceleration amplifies in the run-out zone as time proceeds.
 At at $t = 7$ s the main kink occurs at around $x = 91$ m.
At $t = 8$ s, the situation changes dramatically. There appear two sectors with different drag dynamics. In the mid-rear sector, (80, 90) m, the mass is still moving, and the mass in the frontal sector (90, 115) is preparing to settle down and assumes almost a constant drag coefficient where the peak of the near deposition mass lies at about 90 m.
 The rate of deceleration is high at the front, low at the back and changes accordingly in between with a smooth turn at about 90 m. Consequently, the drag coefficient increases non-linearly from the front to the back in both sectors.
 Now, finally, the mass is about to deposit at $t = 9$ s, where it lies entirely in the run-out zone such that the net driving acceleration is all negative. At $t = 9$ s, as before, still the dynamics can be divided into two sectors. In the frontal sector, (88,~119)~m, the mass is virtually at stand-still with a strong, almost constant drag.
A relatively small portion of it in the back, (86, 88) m, is still moving (decelerating) a bit, but significantly counter acted by the drag.
 Similar dynamics is observed for $t = 10$ s with strong drag coefficient at the still decelerating and nearly depositing mass in the rear, (84, 86) m, whereas, it decelerates heavily in the mid-sector, (86, 93) m, otherwise, in the front, (93, 119)~m, virtually, most of the mass is already deposited with a strong constant drag coefficient.
 There is a very little leisurely moving tail in the back with a sudden jump in drag of an already halted main body in the front.
\\[3mm]
It is observed that for $t = 7 - 9$ s, the inflection pints occur at around 89, 88, 86 m, referring to the changes in dynamics from deceleration to deposition processes.
There is a clear backward propagating deposition waves indicated by the smooth and strong kinks from panel F to J in Fig. \ref{Fig_8}. At $t = 8, 9, 10$ s there appear double kinks indicating the rapid changes of the flow dynamics and the complex process of the mass deposition. However, the near constant drag coefficients downstream of these strong kinks do not mean the drag forces are constant there, because the drag force is given by the product of the drag coefficient with the the momentum transport, i.e., $-\beta h u^2$.
\\[3mm]
Figure \ref{Fig_9} displays the drag coefficients on the inclined (left panels) and the run-out (right panels) sectors when the landslide mass occupies both sectors, corresponding to panels E$-$H in Fig. \ref{Fig_8}. At each time instance, the jumps are large. After transition the drag intensity is decreasing in time because the impact velocity is decreasing.
\\[3mm]
Figure \ref{Fig_8} with Fig. \ref{Fig_6} manifest very impressive performance of the dynamic drag model $\beta$ in (\ref{Eqn_beta10}). Several striking, fully counter intuitive, mechanical phenomena are revealed. Even for small scale simulation, the change in the evolving drag coefficient is more than an order of magnitude in just a couple of seconds and even more so at the flow transition with a strong shift.
All these dynamics are fascinating as they are in line with the mechanism of the drag coefficient, because the physics tells that the drag coefficient is inversely proportional to the flow velocity. This is exactly what is demonstrated in Fig. \ref{Fig_8}.
In contrast to the empirical, user-select, free parameter drag, the advantage of the new, physically described drag model is that, it is well-defined and dynamically evolving.
\\[3mm]
Yet, surprisingly, the dynamical drag coefficients are around the often calibrated values in the literature. This is important, because, our new mechanical drag equation conceivably well reproduces the dynamics of the natural events, but now, with the clear physical basis.

\subsection{Scope of the new analytical drag}

For more direct understanding and the scope of the new drag model, consider three profiles at $t = 4, 6, 10$ s in Fig. \ref{Fig_6}. At $t = 4$ s, the mass lies entirely in the inclined portion of the slope where the net driving acceleration is positive. There are two important aspects to realize here. First, although variable from its front to the rear, for this situation, the drag dynamics is described by (\ref{Eqn_beta3p}), or $\beta^+$ in (\ref{Eqn_beta10}), see panel D in Fig. \ref{Fig_8}. For the profile at $t = 6$ s, as the mass lies both in the inclined and the horizontal sectors of the slope, the drag dynamics for this situation is described by fundamentally different branches of the drag model applied to different flow regimes. As the upper part corresponds to the positive net driving acceleration, (\ref{Eqn_beta3p}), or $\beta^+$ in (\ref{Eqn_beta10}), describes the drag dynamics for this. For the part that lies in the run-out sector of the slope, the net driving acceleration is negative, for which (\ref{Eqn_beta9mb}), or $\beta^-$ in (\ref{Eqn_beta10}), describes the drag dynamics. Yet, both drags are evolving in their own ways in these separate regions, see panel F in Fig. \ref{Fig_8}. However, for the profile at $t = 10$ s, as the mass lies entirely in the run-out region with the negative net driving acceleration, the drag in that region is described by (\ref{Eqn_beta9mb}), or $\beta^-$ in (\ref{Eqn_beta10}), see panel J in Fig. \ref{Fig_8}.
\\[3mm]
Since the drag force is proportional to the square of velocity, it must have strong dynamical reign at the accelerating (early stage of motion) and decelerating (later stage of motion) phases. This is evident in Fig. \ref{Fig_6} with the new analytical drag in efficiently controlling the entire dynamics. This is phenomenal. However, Fig. \ref{Fig_3} – Fig. \ref{Fig_5}, that employ the pre-defined constant drag coefficient lack such control. Perceived from the drag mechanism, this evidently suggests dismissal of the constant drag coefficient approach widely applied in mass flow simulations (Brufau et al., 2000; Chen et al., 2006; Hungr and McDougall, 2009; Christen et al., 2010; Frank et al., 2015; Melo et al., 2018; Salmanidou et al., 2018; Kelfoun, 2021; Shugar et al., 2021; Tayyebi et al., 2022; Meyrat et al., 2023; Singh et al., 2023; Mergili et al., 2025; Sattar et al., 2025).
\\[3mm]
Another important fact about the new dynamic drag model is that it can efficiently and legitimately bring the deposition process to a halt as seen from simulation profiles at time slices $t = 10$ s and $t = 11$ s in Fig. \ref{Fig_6}. The profile for $t = 12$ s virtually coincides with the profile for $t = 11$ s, not shown here. Moreover, after $t = 10$ s, the mass does not diffuse in the rear, rather it stays standing-still. These are useful mechanisms to accurately defining and controlling the final position of the halted mass that other models often suffer to demonstrate.
\\[3mm]
The fronts, rears, run-outs and the peak flow depths, and the propagation speeds are fundamentally different among the  simulations with constant drag coefficients (Fig. \ref{Fig_3} – Fig. \ref{Fig_5}) and between the constant drag coefficients (Fig. \ref{Fig_3} – Fig. \ref{Fig_5}) and the dynamical drag coefficient (Fig. \ref{Fig_6}). The simple question is: which one of Fig. \ref{Fig_3}, Fig. \ref{Fig_4} and Fig. \ref{Fig_5} is physically meaningful? There is no definite answer to this question, simply, because, these simulations are not based on the physically described drag coefficients. The results in Fig. \ref{Fig_6}, on the other hand, is undoubtedly the only physically explained landslide dynamics as explained by the evolutionary, mechanical drag coefficient (\ref{Eqn_beta10}). This conceivably settles down the deliberation on the drag force in landslide dynamics.
\\[3mm]
Both the flow dynamics, the geometrical and morphological structures in deposition are fundamentally different among the simulations with the constant and the dynamical drag coefficients.
In Fig. \ref{Fig_6}, the frontal deposition structure is dominantly controlled by the evolving leading frontal head transiting into the run-out zone, rather than the elongated tail as in Fig. \ref{Fig_3} – Fig. \ref{Fig_5}.
 The front geometry, arrival timing and the velocities are completely different in these two types of simulations. This means, the impact forces are different, arguably better simulated by the new mechanical drag model.
\\[3mm]
The proposed analytical drag coefficient fundamentally controls the state variables, the flow depth and the flow velocity. So, it commands and regulates the intrinsic mechanisms of two of the most important phenomena in mass transport: the erosion-entrainment (Pudasaini, 2025a), and the phase-separation (Pudasaini and Fischer, 2020b). Therefore, the new drag model possesses enormous application potential in properly simulating the complex mass flows.
\\[3mm]
The above results and discussions justify the need for the use of the proposed dynamical drag model in properly simulating mass flow events, because, the drag strongly controls their flow dynamics, run-out and the deposition morphology. This raises a question on the physical legitimacy of the simulation results obtained by random use of the constant viscous drag coefficient without any mechanical foundation (Brufau et al., 2000; Chen et al., 2006; Hungr and McDougall, 2009; Christen et al., 2010; Frank et al., 2015; Melo et al., 2018; Salmanidou et al., 2018; Kelfoun, 2021; Shugar et al., 2021; Tayyebi et al., 2022; Meyrat et al., 2023; Singh et al., 2023; Mergili et al., 2025; Sattar et al., 2025). However, the simulation results presented here with the analytical, dynamical drag is subject to scrutiny with the laboratory and field data.

\section{Summary}

As an important energy dissipation mechanism, drag plays a vital role in controlling the dynamics and deposition characteristics of mass flows including avalanches, landslides and debris flows. Yet, the existing drag models are exclusively empirical and lack the physical foundation. Here, we developed a simple analytical model for drag coefficient in landslide. The model is based on the acceleration number, a dimensionless number, as the ratio between the flow acceleration and the net driving acceleration describing the overall dynamical aspect of the landslide. This innovative mechanical model comprehensively explains the drag. It includes the basic ingredients of the flow physics and dynamics in an unified way. The evolutionary drag coefficient adjusts automatically during the landslide motion as it evolves together with the complex flow dynamics.
The drag coefficient appears to be the measure of landslide (immobility, or) energy inefficiency {as it incorporates the viscosity of the ambient fluid}.
This addresses the long-standing question in mechanically explaining the drag force in landslide dynamics, defining a new perspective of the drag. Simulation results manifest the functionality, mechanical strength, superiority and the scope of the proposed analytical drag model against the empirical drag models as it displays some essential behaviour of the frictional debris mass flows. The engineers and practitioners may find the new drag model useful in properly simulating catastrophic mass transport events in mountain flanks and vallies.

\section*{Acknowledgements}

The author acknowledges the financial support from the German Research Foundation (DFG) through the research project: Landslide mobility with erosion: Proof-of-concept and application - Part I: Modeling, Simulation \& Validation; Project number 522097187.

\end{document}